\documentclass[useAMS,usenatbib]{mn2e}

\usepackage{graphicx}
\usepackage{journals}
\usepackage{url}
\usepackage{multirow}

\usepackage{amsmath}

\usepackage{natbib}

\title[Search for photon bubble oscillations in V0332+53]{Search for photon bubble oscillations in V0332+53}
\author[Revnivtsev et al.]{Mikhail G. Revnivtsev$^{1}$\thanks{E-mail:revnivtsev@iki.rssi.ru}, Sergey V. Molkov $^{1}$, Mikhail N. Pavlinsky$^{1}$
\\
$^{1}$ Space Research Institute, Russian Academy of Sciences, Profsoyuznaya 84/32, 117997 Moscow, Russia\\
}
\begin{document}

\date{}

\pagerange{\pageref{firstpage}--\pageref{lastpage}} \pubyear{2015}

\maketitle

\label{firstpage}

\begin{abstract}
We report results of our search for fast oscillations in lightcurve of one of the brightest accretion powered pulsars on the sky V0332+53 with the help of data of the PCA spectrometer of the RXTE observatory. In course of this search we have carefully explored complications appearing if one uses only sub-bands of the total bandpass of the PCA spectrometer. We show that lightcurves collected in the soft sub-band of the PCA spectrometer contains an additional instrumental noise, lightcurves of harder sub-bands lack some fraction of the anticipated Poisson noise. We show that this noise is caused by a cross-talk of energy bands, which lasts up to $\sim200\mu$sec. One hypothesis is that these effects are caused by temporarily drop of the PCA detector gain after any occurred event due to slowly moving ions in the detector volume. In order to avoid this effect we searched for fast oscillations in flux of V0332+53 only in the total bandpass of the PCA spectrometer 2-60 keV. We have not detected any quasi-periodic oscillations in lightcurve of the source with an upper limit at the level of 0.5\% in the Fourier frequency range 200-1500 Hz.
\end{abstract}


\section{Introduction}
Binary systems with a compact object are among the brightest sources in galaxies, in particular in the X-ray energy range. Their bolometric luminosity is powered by accretion onto relativistic objects like neutron star (NS) or black hole (BH). Luminosity of black hole accretors is produced by accretion disks \citep{ss73,pringle72}, while neutron star accretors have additional structures on their surfaces. In case of non magnetic neutron stars it is a boundary/spreading layer,  in which rapidly rotating matter of the accretion disk decelerate to neutron star rotational velocity \cite[e.g.][]{hoshi84,sunyaev86,inogamov99}. If the neutron star is strongly magnetized, then the accretion flow can not continue undisturbed till  the neutron star and is channelled to neutron star magnetic poles \citep{lamb73,davidson73}. 

In case of actively accreting neutron stars the falling matter is stopped near the NS surface due to pressure of outgoing radiation \citep{basko76,braun84,lyubarskii88}. In a set of papers it was shown that the settling of the infalling flow to the NS surface might be unstable with respect to presence of so-called photon bubbles -- the regions with high radiation pressure act like light fluid with respect to heavy fluid of infalling plasma \cite[see e.g.][]{arons92,klein96a}. It was predicted that these oscillations might be visible in time series of X-ray luminosity of such sources. Depending on the area of the accretion column footprint estimates of the characteristic timescales varied from hundreds of Hz to several kHz \citep{klein96a,jernigan00}. 

Launch of the orbital observatory Rossi XTE \citep{bradt93} have provided a tool to search for these predicted features. The PCA spectrometer of RXTE observatory had the largest collecting area of all previous X-ray orbital instruments (it was switched off on Jan.5, 2012) and indeed allowed to discover a lot of new fast phenomena \cite[see e.g.][]{vdk06}. Among them there were claims about detection of photon bubbles oscillation \citep{klein96b,jernigan00}. 

In this paper we are trying to obtain the best possible constrains on presence of a rapidly (at Fourier frequency scales around kHz) variable noise component in lightcurves of accretion powered X-ray pulsars. As an example we study one of the brightest source of this type V0332+53.

\section{Data analysis}

In our search we use data of the PCA spectrometer \citep{jahoda06} of the RXTE observatory \citep{bradt93}

The most straightforward way to search for quasi-periodic oscillations is to study power density spectra of original lightcurves of X-ray sources, calculated as square of amplitude of their Fourier transform \citep{leahy83}:

\begin{equation}
P_j=\frac{2|a_j|^2}{N_\gamma}
\end{equation}

\noindent
where, $|a_j|$ -- amplitude of the Fourier transform of the curve at Fourier frequency $j/T$, $N_{\gamma}$ -- number of photons within the transformed lightcurve segment of length $T$.
Due to limited number of photons from X-ray sources, variability of a signal at the highest Fourier frequencies is always dominated by Poisson noise. In the abovementioned normalization the level of Poisson noise to be seen by an ideal instrument is 2.0 
\citep{leahy83}

\subsection{Power spectra from PCA data and effects of deadtime}

In the case of a real instrument, this counting noise of photons is  modified by a number of instrumental effects. PCA is a gas proportional counter, consisting of 5 independent identical units (Proportional Counter Units - PCUs)  each with its own electronics. All events, which trigger the low level discriminator initiate a sequence of analog-to-digital conversions, during which the detector is busy -- the detector is deadtime-locked. All units count X-ray photons, but also count a large flux of background events, created by charged particles. Charged particles typically create events at more than one set of the PCU anodes and thus a large fraction of them can be vetoed using anti-coincidence logic. This explicitly means that the "good counts"  is only a fraction of total counts, measured by instrument. In case of bright sources the fraction of good counts is large but still less than unity.

In addition to simple background events PCUs also have so called Very Large Events -- events, which deposit more than $\ga 75$ keV of energy into the PCU volume. These events saturate PCU analogue amplifiers and in order to resume nominal operation of the detector, the whole PCU is intentionally disabled for some pre-set time period. Typical settings, which can be found in scientific data is {\tt dsVle=1} and {\tt dsVle=2}. Exact values of these VLE dead-times during the flight of RXTE/PCA were measured by different methods to be approximately 50-70 $\mu$sec and 140-170 $\mu$sec correspondingly \citep{zhang95,jernigan00,revnivtsev00,wei06,jahoda06}.

Effects of modification of pure Poisson noise in data of RXTE/PCA were identified in a number of works \cite[see e.g.][]{zhang95,zhang96,morgan97,jernigan00,revnivtsev00,wei06}.
It was shown that the main predicted effects of timing logic of PCA detector chains on Poisson noise of counts are indeed seen in real data, but it is exceedingly hard to calculate exact corrections from the first principles and some recorded housekeeping count rates. Thus, in some works it was adopted to use an analytic model of deadtime-modified Poisson noise and fit its parameters to real power spectra \citep{jernigan00,sunyaev00,revnivtsev00}. This approach uses assumption that at high frequencies (typically above kHz) there is no intrinsic variability of flux of real sources and all recorded variability is due to modified Poisson process.

\begin{figure}
\includegraphics[width=\columnwidth,viewport=40 180 570 420,clip]{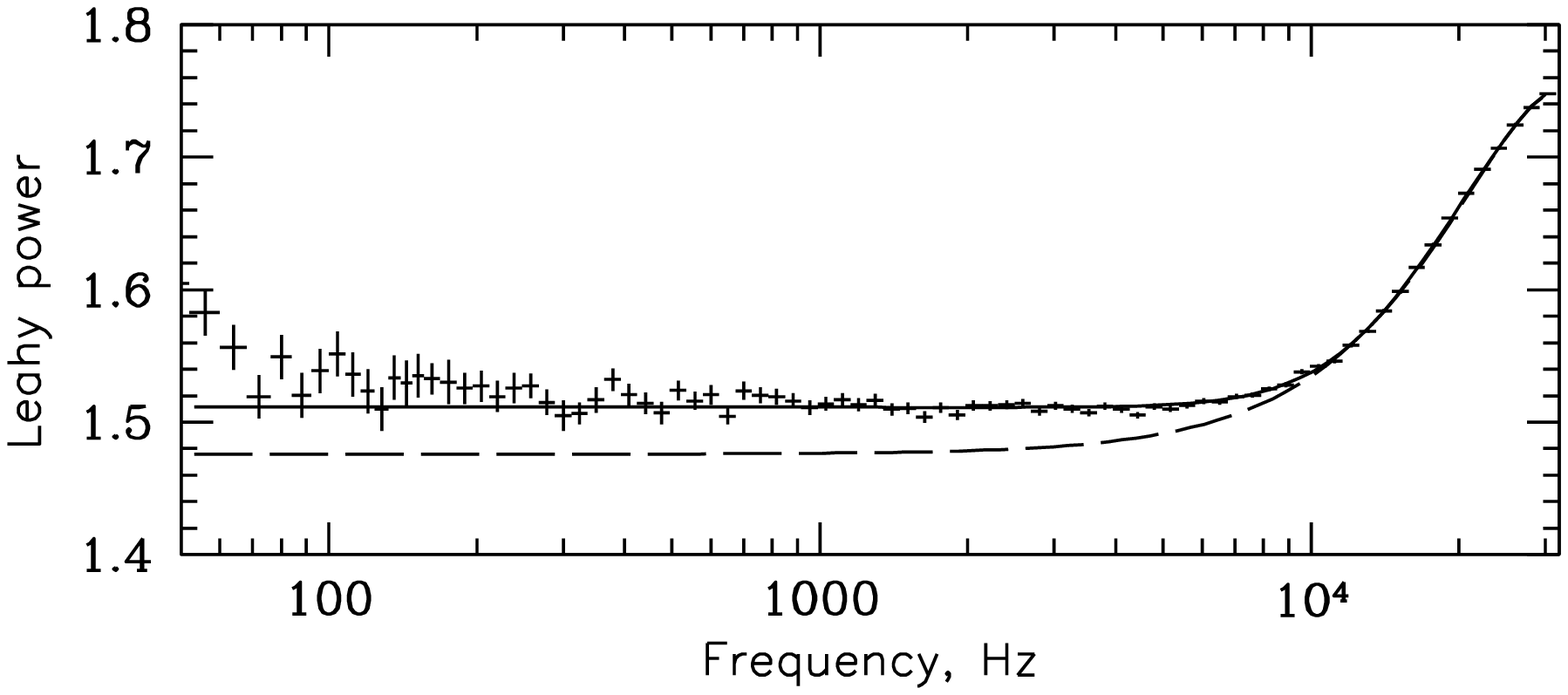}
\caption{Power spectrum of Sco X-1 in total energy band of RXTE/PCA from Obd-ID 10059-01-03-00 (Feb.19, 1996) along with the best fit model of the deadtime-modified Poisson noise. The VLE deadtime setting during this observation is {\tt dsVle=1}, for which we adopted value 76 $\mu$sec. Dashed curve shows the model of deadtime modified Poisson noise without contribution of the VLE component.}
\label{scox1_broad}
\end{figure}

It was shown that modification of the Poisson noise in total PCA bandpass due to simple deadtime effect and effect of occasional switch-offs due to Very Large Events can be adequately described by formulas \cite[see e.g.][]{jahoda06}:

\begin{equation}
\label{eqn2}
P(f)=2-P_1-P_2\cos\left({2 \pi t_b f}\right)+P_3\left({\sin \pi t_{\rm vle} f\over{\pi t_{\rm vle} f}}\right)^2
\end{equation}
$$
P_1=4r_0 t_d\left(1-{t_d\over{2t_b}}\right)
$$
$$
P_2=2r_0t_d {N-1\over{N}}\left({t_d\over{t_b}}\right)
$$

$$
P_3=2 r_0 r_{\rm vle} t_{\rm vle}^2
$$
\noindent 
where $t_d\sim 8-10~\mu$sec -- deadtime of the instrument per event, $t_{\rm vle}$ -- deadtime of Very Large Event, $t_b$ -- bin time of the lightcurve, $r_0$ -- average count rate in the lightcurve, $N$ -- number of frequency bins in the Fourier transform. In these formulas all count rates are written for single detector. Example of application of these formulas to power spectrum of the brightest source on X-ray sky Sco X-1 is shown in Fig.\ref{scox1_broad}. For this plot we have used observation ObsID 10059-01-03-00, count rate of the source is $\sim77.5$ kcnts/sec/5PCU, exposure time 1.04 ksec. We adopted here $t_{\rm vle}=76~\mu$sec \citep{revnivtsev00}. It is seen that quality of this analytic fit is very good: $\chi^2=62.2$ for 55 degrees of freedom for power density values at Fourier frequencies above 400 Hz.

\subsection{Modifications of power spectra in PCA energy sub-bands}

In works of \cite{revnivtsev00,wei06} it was noted that apart from above-mentioned modifications of Poisson noise in real data some additional modifications are revealed in power density spectra calculated from data in sub-bands of RXTE/PCA. In particular, the most striking feature is an appearance of an additional noise at frequencies up to 1-1.5 kHz in lightcurves, constructed from data at lowest energy ranges of PCA (e.g. channel range 0-13 or 0-17 out of 256 channels of PCA).

As an illustration of this effect we present power spectra of the brighest X-ray source on the sky Sco X-1 from a set of observations ID 96443-03-01-\{00,...,13\}, 96443-03-02-00 (time resolution $2^{-14}$ sec $\sim61\mu$sec), total exposure time $\sim9.5$ ksec.  We have caluclated power spectra for three energy bands: channels 0-17 ($<7$keV, count rate $\sim$75 kcnts/sec/5PCU), 18-35 ($\sim$7-15 keV, count rate $\sim$28 kcnts/sec/5PCU) and 0-249 (total energy range of PCA 2-60keV, count rate $\sim103.7$ kcnts/sec/5PCU). The VLE deatime setting during these observations is the same ({\tt dsVle=1}) as during observation, shown in Fig.\ref{scox1_broad}. 

It is seen that while the total band (channels 0-249) power density spectrum is indeed similar to that on Fig.\ref{scox1_broad} (apart from different time binning and somewhat larger source count rate), the power density spectra from sub-bands (channels 0-17, channels 18-35) are strongly different from that.

\begin{figure}
\includegraphics[width=\columnwidth,viewport=1 200 570 710, clip]{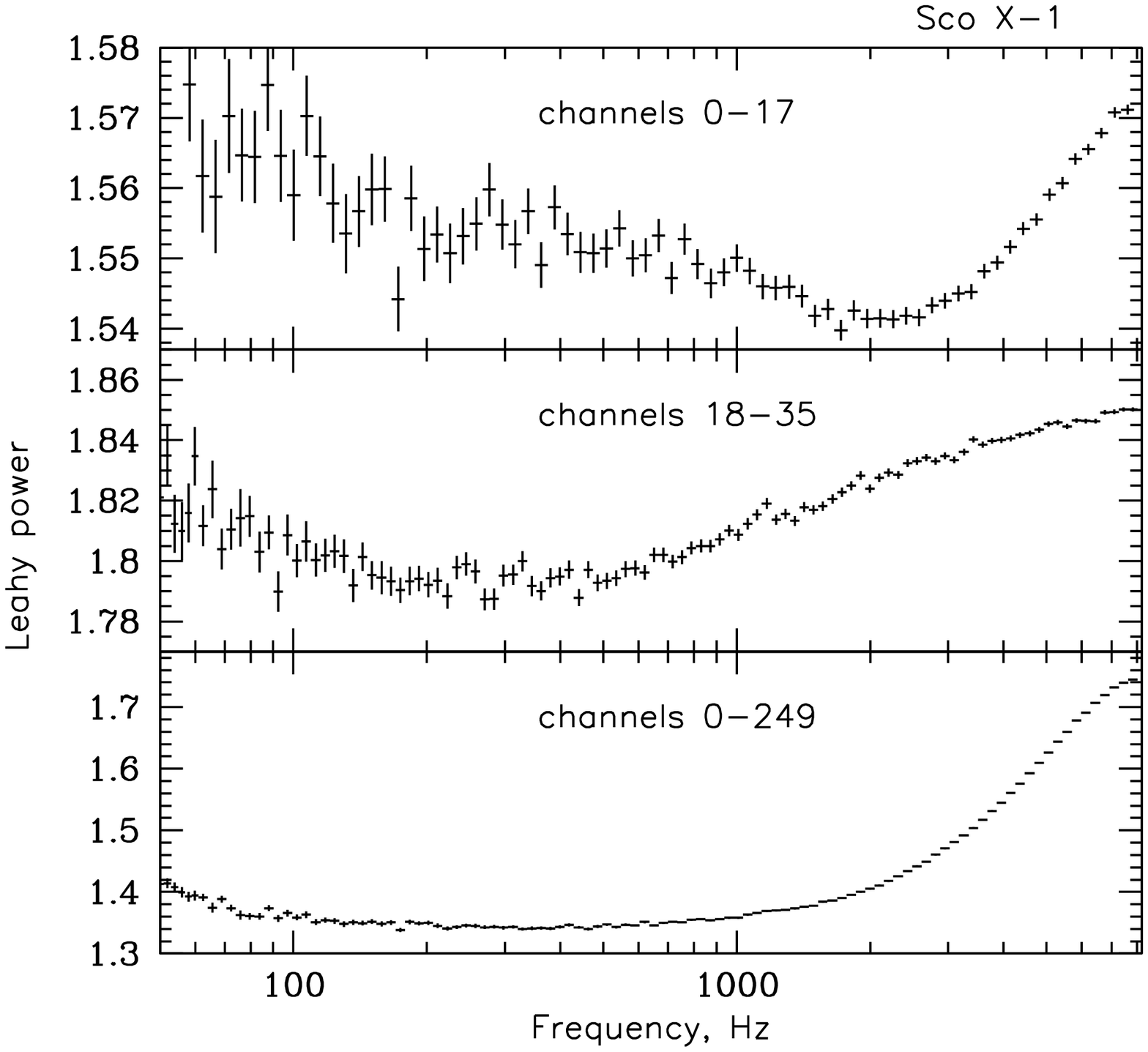}
\caption{Power spectra of lightcurves of Sco X-1 from observation IDs 96443-03-.. in three energy bands: 0-17, 18-35 and 0-249 channels.}
\label{scox1_chans}
\end{figure}

\begin{figure}
\includegraphics[width=\columnwidth,viewport=10 250 570 710, clip]{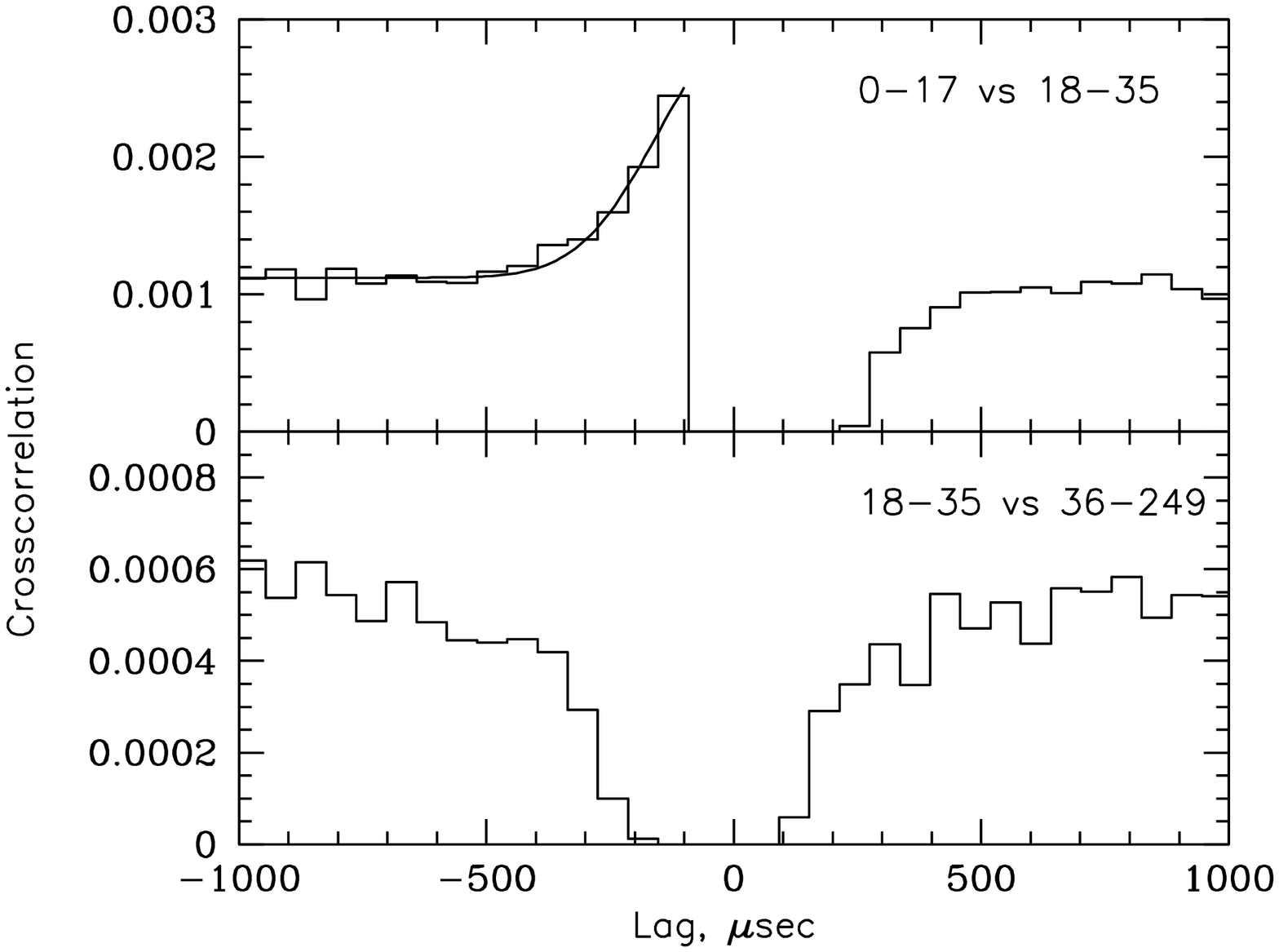}
\caption{Cross-correlations between lightcurves of Sco X-1 in channels 0-17, 18-35 and 36-249 (corresponding to energies $<7$ keV, $\sim$7--15 keV and 15--60 keV).}
\label{cc_scox1}
\end{figure}

It is likely that the origin of these peculiarities is related to physics of detection of X-ray photons in a gas proportional counter.
In work of \cite{wei06} it was proposed that this effect might be caused by temporarily drop of the detector gain over some part of the detector volume immediately after any occurred event.  All events depositing energy into detector volume, create a cloud of electron-ion pairs. Electrons drift to the nearest anode and there create an avalanche. Electrons born in this avalanche are rapidly absorbed by the anode, but much slower ions drift to the respective cathode. During their drift time they can screen the electric field for subsequent events and thus effectively reduce the detector gain. Events, occurred during time periods, when ions from the previous event are drifting towards the cathode, will have slightly lower pulse heights/ascribed energies.
Thus, any event, which occurred in the detector volume causes temporarily small drop of the gain in this volume. 

It means that events, occurred near the boundary between softer and harder bands will be able to switch from harder band to softer band if they will come during this "floating boundary" time period. Temporarily drop of the detector gain squeeze the energy scale of the instrument. Therefore only hard energy bands can loose counts. The softest energy band does not loose counts, but rather gather counts from neighbouring harder energy band.

In order to check presence of this effect in real data we have calculated cross correlations between light curves of Sco X-1 in three non-overlapping energy bands (the same data as used for Fig.\ref{scox1_chans}). If there would be no "cross-talks" between counts in these energy channels apart from ordinary dead-time effect due to detector electronics we should not expect to see any effects on delays $\tau>|t_d|\sim 8-10\mu$sec. In fact we do see very strong correlation of these signals up to delays $<300~\mu$sec (Fig.\ref{cc_scox1}), where intrinsic variability of the source is absent. 

Top panel of Fig.\ref{cc_scox1} shows cross-correlation between energy bands determined by PCA channels 0-17 and 18-35, bottom panel - between 18-35 and 36-249. We see that after any recorded event (zero lag) there is a gap at positive delays during $\sim300~\mu$sec, i.e. the hard band count rate is lower than it should be. We see these missing events at negative delays -- they switched to be in the soft band. The effective timescale for this "floating boundary" window is approximately $\sim200-300\mu$sec. The amplitude of the peak at negative delays (and amplitude of the gap at positive delays) should depend on the fraction of counts, which appear as additional to the original counts in the soft energy band for the "extended ion drift time" period. 

Crosscorrelation of light curves in harder energy bands (Fig.\ref{cc_scox1}, lower panel) looks a bit different. We also see a gap at positive delays -- switching of fraction of harder band photons into softer band, but at the same time we also see a gap at negative delays. This happens because during finite time period after any event the count rate in this band (in our case channels 18-35) was increased by some fraction of counts from harder band (channels 36-249), but at the same time much larger fraction of counts moved from this band to even softer band (channels 0-17), thus in total this energy band still loose some fraction of its counts.

The characteristic time scale of the noted effective change of the energy band boundaries can be estimated from cross-correlations: a gaussian fit gives approximately $t_{\rm drift}\sim 160-170\mu$sec (see Fig.\ref{cc_scox1}). It should be compared to the time of drift of ions in PCU volume. 
Xenon ion mobility is approximately 0.8 cm$^2$ sec$^{-1}$ V$^{-1}$ (e.g. Tables of Physical \& Chemical Constants, Kaye \& Laby Online, www.kayelaby.npl.co.uk ). For the applied voltage $\sim2.5$ kV/cm  the ions drift velocities is $\sim 2$ cm/msec. The distance between anodes and cathodes is approximately 0.6-0.7 cm (the total thicknes of PCU layers is approximately 1.3 cm, \citealt{jahoda06}),  therefore the drift time is approximately $\sim 300\mu$sec. This estimate is broadly consistent with the timescale, which we inferred from analysis of cross-correlations.

Therefore we do see that the PCA scheme of readout of events into separate energy bands have relatively long relaxation time and thus lightcurves of sources, recorded in these energy bands are mutually distorted.

\subsection{Simulations}

We have simulated the abovementioned effect.

\begin{figure}
\includegraphics[width=\columnwidth,bb=22 183 586 512,clip]{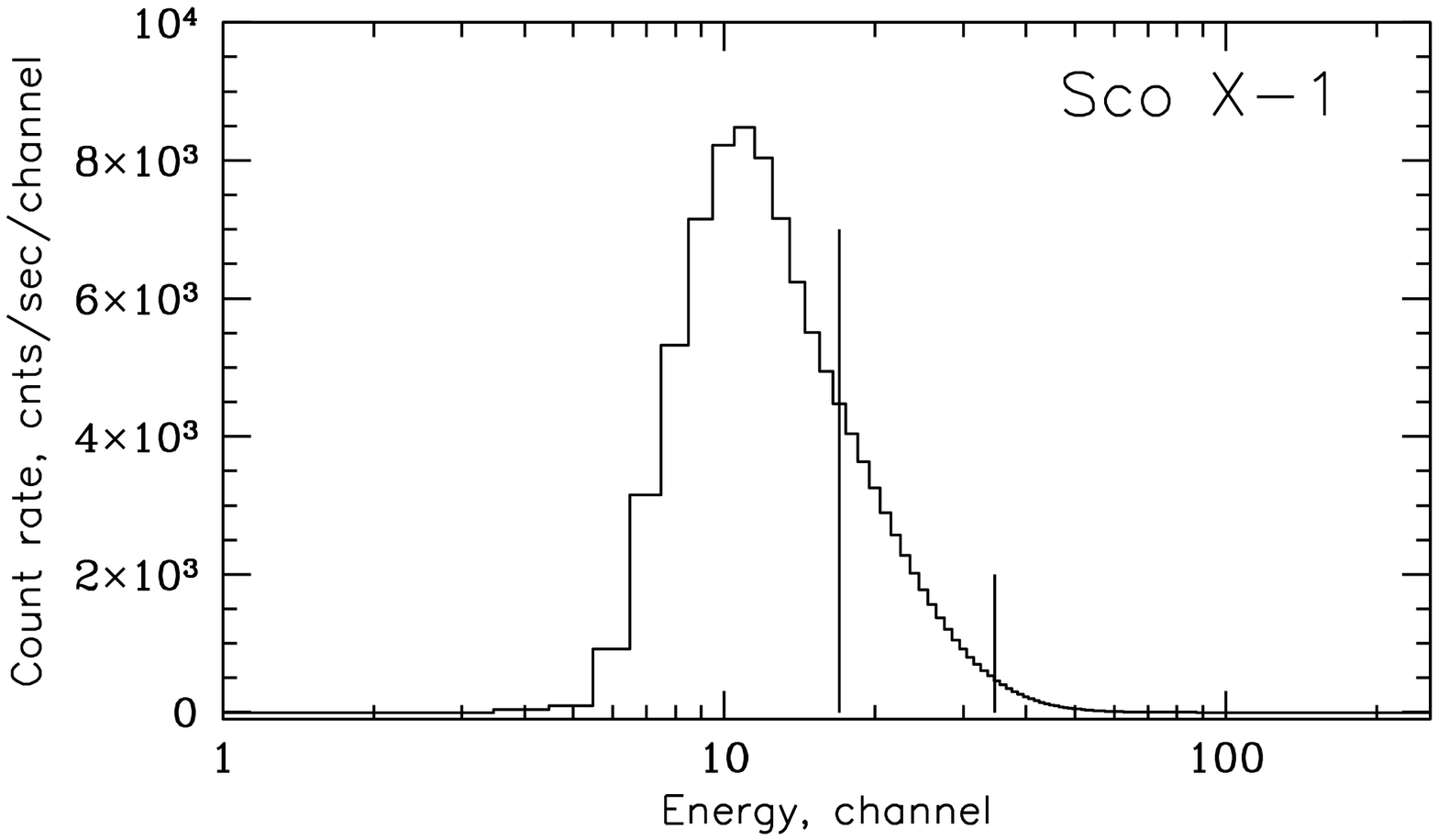}
\caption{Spectral energy distribution of Sco X-1 used in simulation (recorded by RXTE/PCA in observations ObsIds 96443-03-...). Vertical lines denote boundaries between energy bands (0-17, 18-35, 36-249).}
\label{scox1_spec}
\end{figure}

We have simulated pure Poisson noise of photons and selected only those, which were separated by more than non-paralizable deadtime $t_d=10\mu$sec. For all events we have simulated the energy (channel), according to spectral energy distribution of Sco X-1, recorded in ObsIDs 96443-03-... We have adopted the intrinsic source count rate 20 kcnts/sec/PCU in total energy range (which gives 100 ksec/sec/5PCU, PCA spectrometer has 5 independent detector units). {

\sl We have not simulated the VLE deadtime effect}. 

In order to imitate the temporarily shift of boundaries between energy bands we have assumed that any incoming event with some predefined probability to cause a change of the energy scale of the instrument (gain). We assumed that the depth of the gain drop depends on energy of the event. The energy scale (gain) dependence on time since that event ($dt$) was assumed to be:
$$
G(dt)=1-0.15 \left[{E\over{7 \textrm{keV}}}\right]\exp\left[-{1\over{2}}\left({dt\over{t_{\rm drift}}}\right)^2\right]
$$
, here the ion drift time $t_{\rm drift}=170\mu$sec, $dt$ -- time between current event and that caused the gain drop. The final energy channel of the event is $ch_{f}=ch_{i} G$. The resulted event list was split into three energy bands 0-17, 18-35 and 36-249.

\begin{figure}
\includegraphics[width=\columnwidth,bb=26 180  565 710,clip]{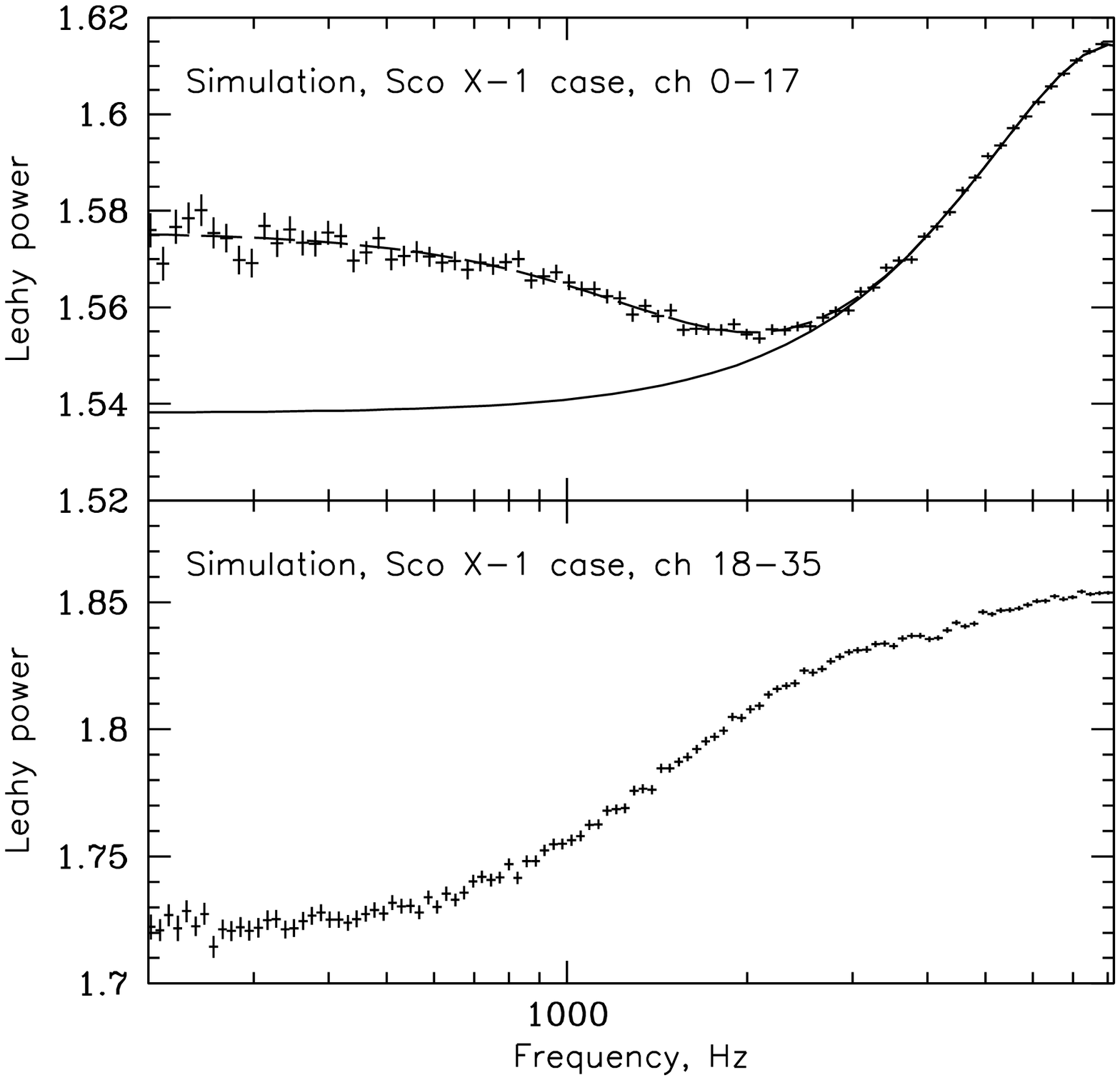}
\caption{Power spectrum of simulated "Sco X-1" lightcurve in channel range 0-17. Solid curve shows the adopted model without "floating boundary" effect. Dashed curve illustrates an additional power introduced by a "floating boundary". In this case it is approximated by a zero-centered gaussian with characteristic width $f_0=1043\pm20$ Hz}
\label{power_sim_ch0_17}
\end{figure}

\begin{figure}
\includegraphics[width=\columnwidth,viewport=10 250 570 710, clip]{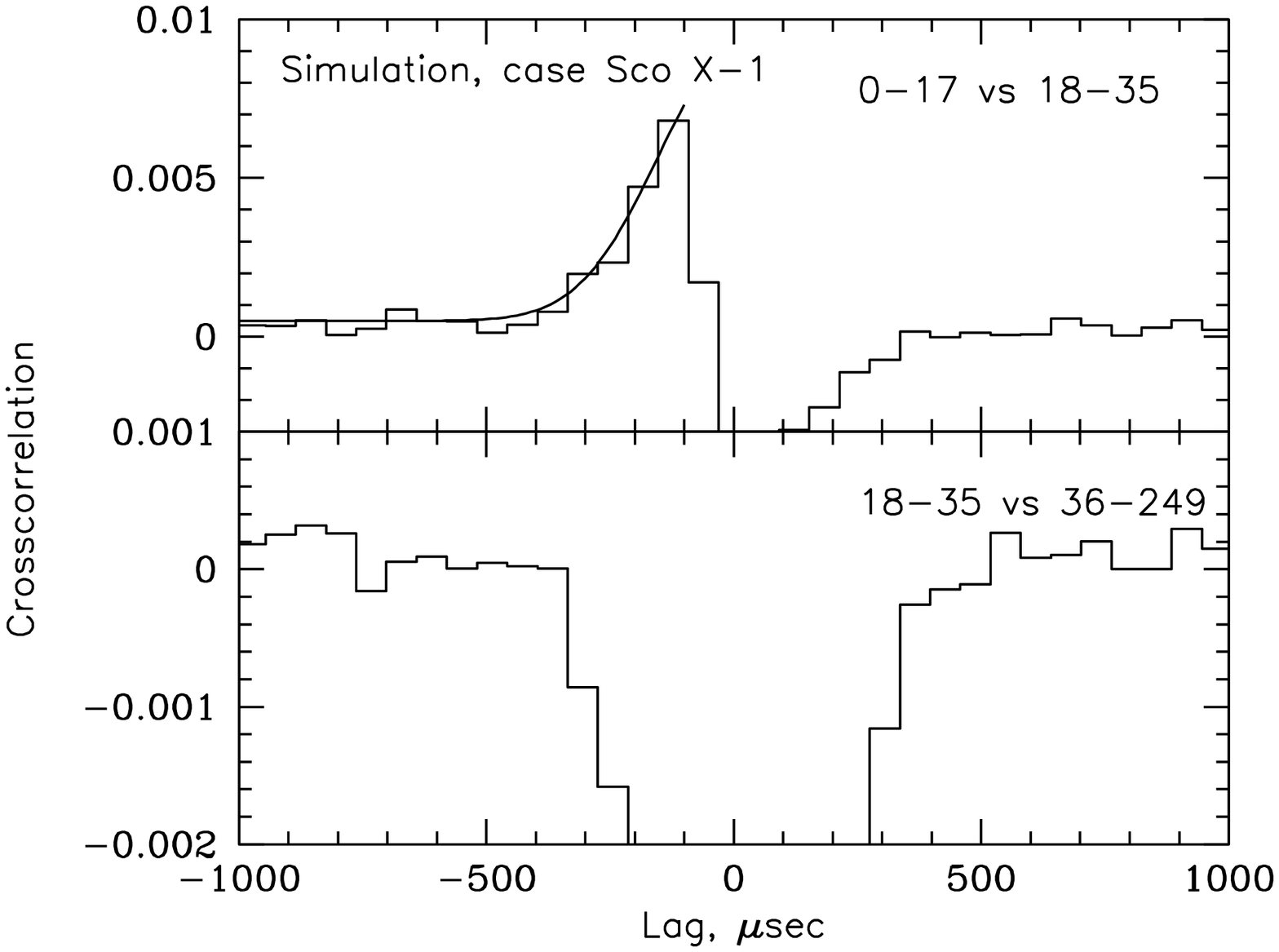}
\caption{Cross-correlations between simulated lightcurves (case adopted to Sco X-1) in channels 0-17, 18-35 and 36-249.}
\label{crosscor_sim}
\end{figure}

Power spectra of lightcurves (total exposure time 20 ksec) obtained in this way in channels 0-17 and 18-35 are shown in Fig.\ref{power_sim_ch0_17}.
Crosscorrelations of obtained lightcurves are shown in Fig.\ref{crosscor_sim}.

Additional component that one can see at power spectrum collected in channel range 0-17 can be approximated by a zero-centered gaussian:

\begin{equation}
\label{eqn3}
P_4 \propto \exp\left[-{1\over{2}}\left({f\over{f_{0}}}\right)^2\right]
\end{equation}
\noindent
with characteristic width $f_0=1043\pm20$ Hz.

\begin{figure}
\includegraphics[width=\columnwidth,viewport=10 150 570 513, clip]{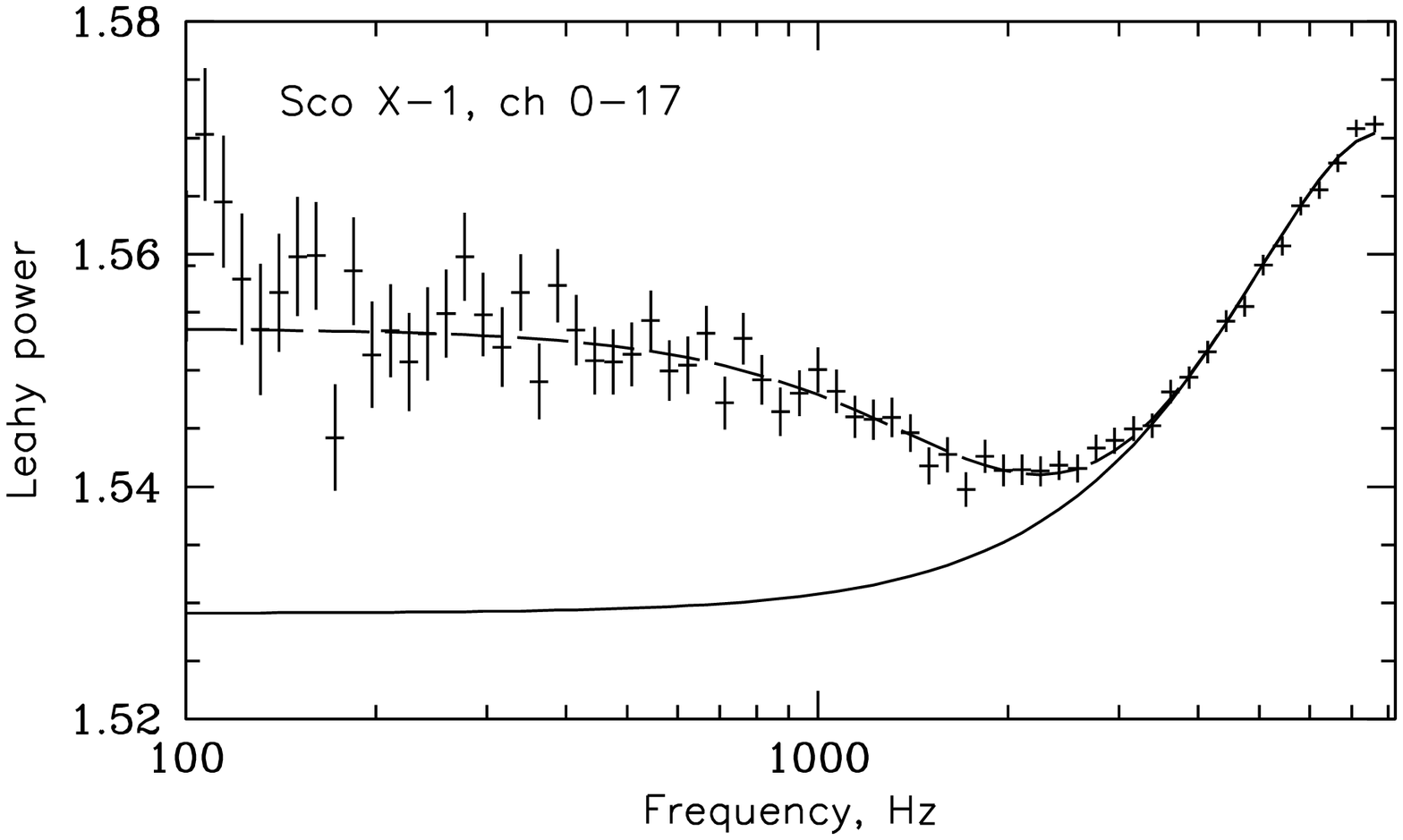}
\caption{Power spectrum of the Sco X-1 lightcurve in the channel band 0-17 ($<7$ keV). Solid lines shows the best fit model to the modified Poisson noise in the lightcurve. Contribution of noise due to "floating boundary effect" is illustrated by a dashed curve, it is approximated by a zero-centered gaussian with characteritic width $f_0\sim 1060$ Hz.}
\label{power_ch0_17}
\end{figure}

Example of approximation of the power spectrum of Sco X-1 soft band lightcurve (channels 0-17) by models of deadtime modified Poisson noise, i.e using formulas (\ref{eqn2}, including effect of VLE deadtime) and effect of the "floating boundary noise" (\ref{eqn3}), is shown in Fig.\ref{power_ch0_17}. The effect of "floating boundary" can be approximated by a zero-centerd gaussian with $f_0=1060\pm70$ Hz.

\subsection{Cen X-3}

One of the results, which might be affected by the abovementioned instrumental feature is the detection of a continuum noise around kHz in bright X-ray pulsar Cen X-3. In work of \cite{jernigan00} it was shown that power spectrum of the Cen X-3 lightcurve, recorded by PCA in the soft energy band (channels 0-17, which corresponds to energies $<$7 keV) demonstrates broad kilohertz continuum (possibly with two quasi-periodic oscillations). Shape of this noise component was estimated by fitting formulas (\ref{eqn2}) to data above $\sim$2 kHz. Formulas (\ref{eqn2}) were shown to be quite adequate for the total energy band of the RXTE/PCA, but as we now see not fully adequate for the PCA sub-bands.

\begin{figure}
\includegraphics[width=\columnwidth,bb=17 189 572 436,clip]{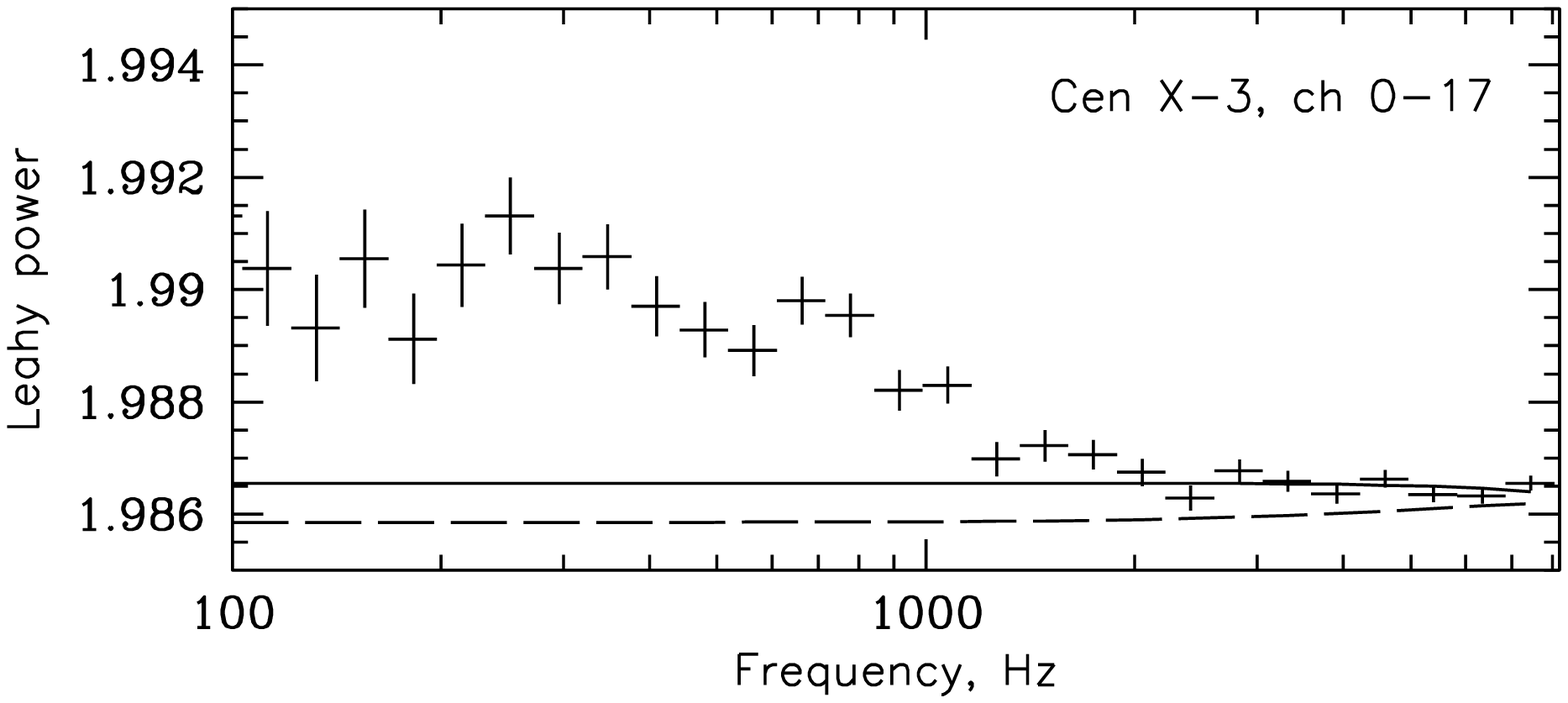}
\caption{Power spectrum of Cen X-3 collected in channel range 0-17. Solid curve is the best fit approximation of a model from Eqns. (\ref{eqn2}) to the data at Fourier frequencies above 2 kHz. Dashed curve is a contribution of non VLE-deadtime components.}
\label{cenx3power}
\end{figure}

In order to demonstrate the importance of effect of "floating boundary noise" for case of Cen X-3 we have repeated the analysis of \cite{jernigan00}. We have selected the same dataset (ObsIds 20104-01-01-..., taken on Feb 28.-Mar.3, 1997), and used only time periods when the source was not eclipsed (count rate in the channel range 0-17 is more than 700 cnts/sec/5PCU). This gives us an exposure time $\sim208$ ksec. The power spectrum in the channel range 0-17 (Fig.\ref{cenx3power}) was constructed over 16 sec segments  ($2^{18}$ time bins with $t_b\approx 61\mu$sec) and then averaged and binned into logarithmically spaced bins.

\begin{figure}
\includegraphics[width=\columnwidth,bb=22 183 586 512,clip]{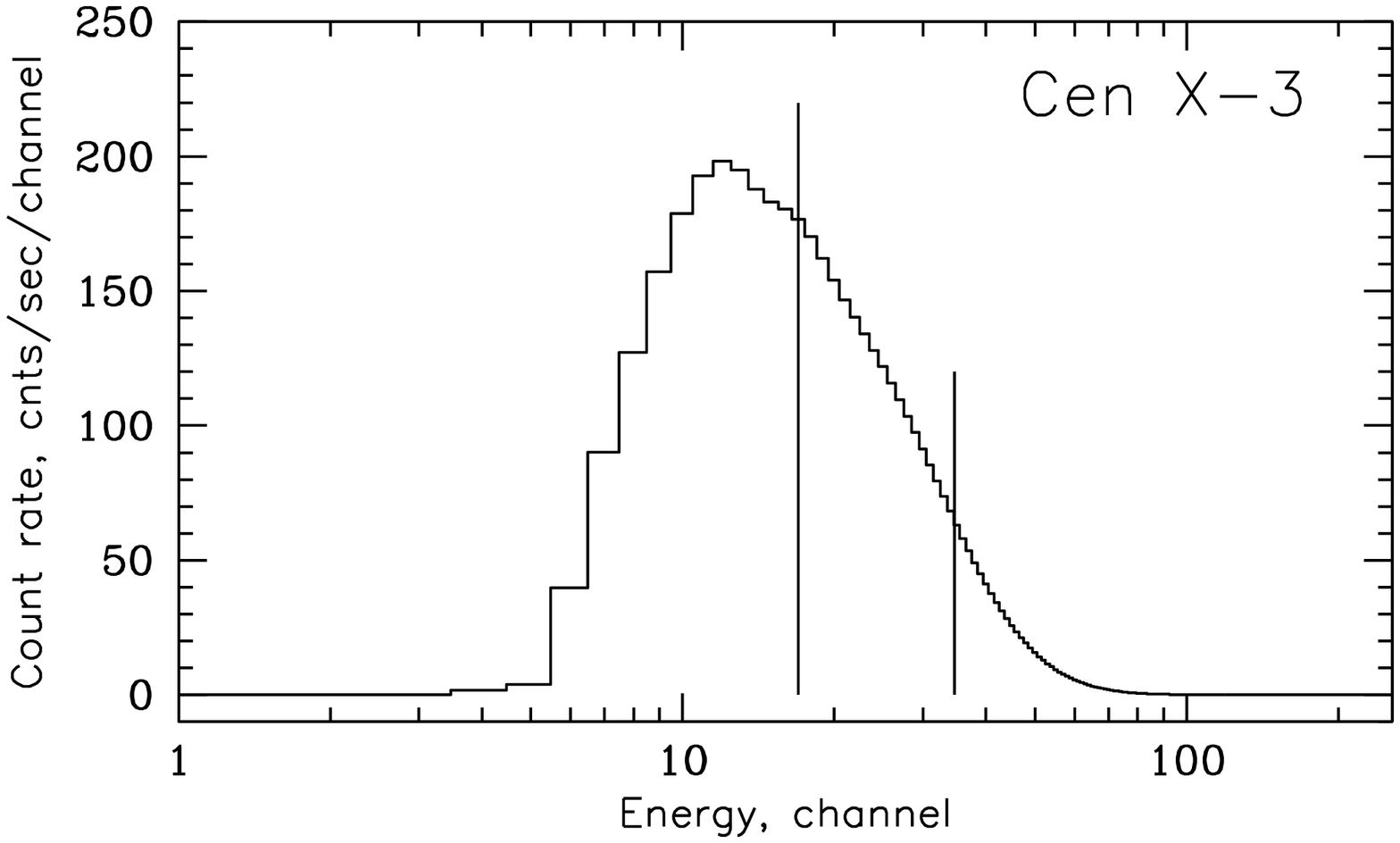}
\caption{Spectral energy distribution of Cen X-3 used in simulation. Vertical lines denote boundaries between energy bands (0-17, 18-35, 36-249).}
\label{cenx3ensp}
\end{figure}

\begin{figure}
\includegraphics[width=\columnwidth,bb=26 180  565 509,clip]{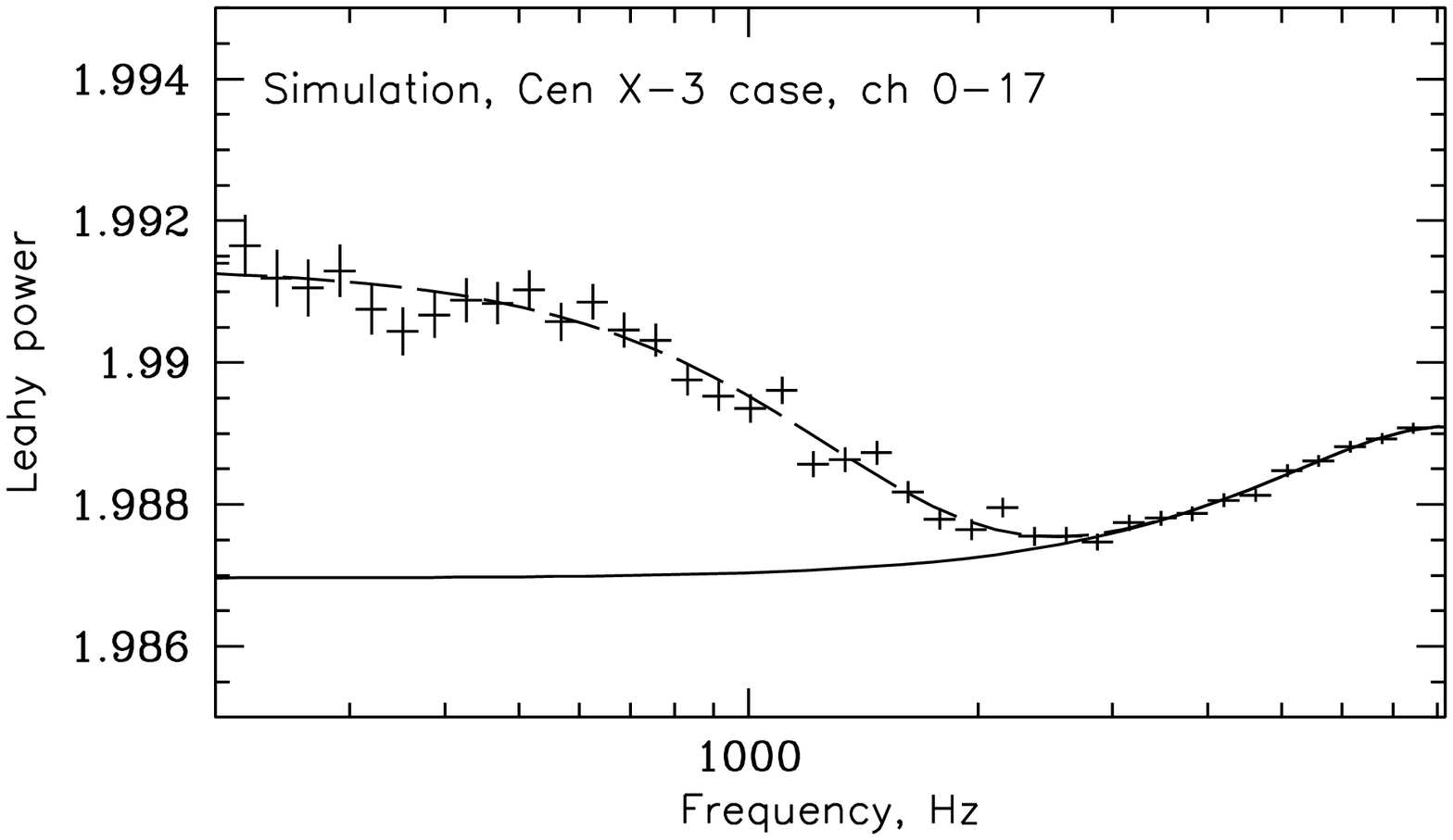}
\caption{Power spectrum of simulated lightcurve of a source with energy spectrum of Cen X-3. Contribution of "floating boundary" noise is shown by dashed curve. It is approximated by a zero-centered gaussian with effective width $f_0=940\pm30$ Hz.}
\label{cenx3sim}
\end{figure}

The energy spectrum of Cen X-3 is strongly different from that of Sco X-1. As the number of counts which are able to switch between energy bands depends on the shape of the spectrum we need to make another simulation of the effect of the ion-drift time on power spectra of the source lightcurves. 

The energy spectrum of the source was taken from real data (see Fig.\ref{cenx3ensp}). Power spectrum of the simulated lightcurve in the channel range 0-17 is presented in Fig.\ref{cenx3sim}. Simulation was performed with the same setup as for Sco X-1 but with lower count rate. The adopted intrinsic count rate of the source in the total energy band is $r_{\rm Cen X-3}=770$ cnts/sec/PCU.

The resulted power spectrum of a simulated lightcurve (the channel range 0-17, total simulated exposure time $\sim$518 ksec) is shown in Fig.\ref{cenx3sim}.

It is seen that the "floating boundary" effect creates an increase of variability power exactly at the place where it was seen by \cite{jernigan00}. Therefore we should conclude that the increased noise in signal from Cen X-3 at frequencies above 100 Hz can be interpreted in terms of the described instrumental effect.

As an additional (but not decisive) argument in favour of artificial nature of this signal one can mention that the signal (broad kHz continuum) was not detected in higher energy band ($\sim7-15$ keV), while the average count rate in this band is similar to that in the soft band and typically the amplitude of all QPO-like features does not decrease with energy.

\section{Search for photon bubble oscillations in V0332+53}

Accreting X-ray pulsar in the binary system V0332+53 \citep{terrel83} is one of the brightest sources of this class in our Galaxy. The RXTE observatory have performed a number of observations of this source \cite[e.g.][]{tsygankov06} with total exposure more than 500 ksec. 

In order to obtain the best possible sensitivity to fast oscillations in lightcurve of the source we have selected only time periods, when the total count rate recorded from the source by PCA is larger than 4000 cnts/s/5PCU and analysed only data with time resolution not worse than $\sim122~\mu$sec. In order to avoid problems with the PCA energy sub-bands (described above) we have used combined data over all PCA energy channels (0-249).

\begin{figure}
\includegraphics[width=\columnwidth,viewport=10 150 570 710, clip]{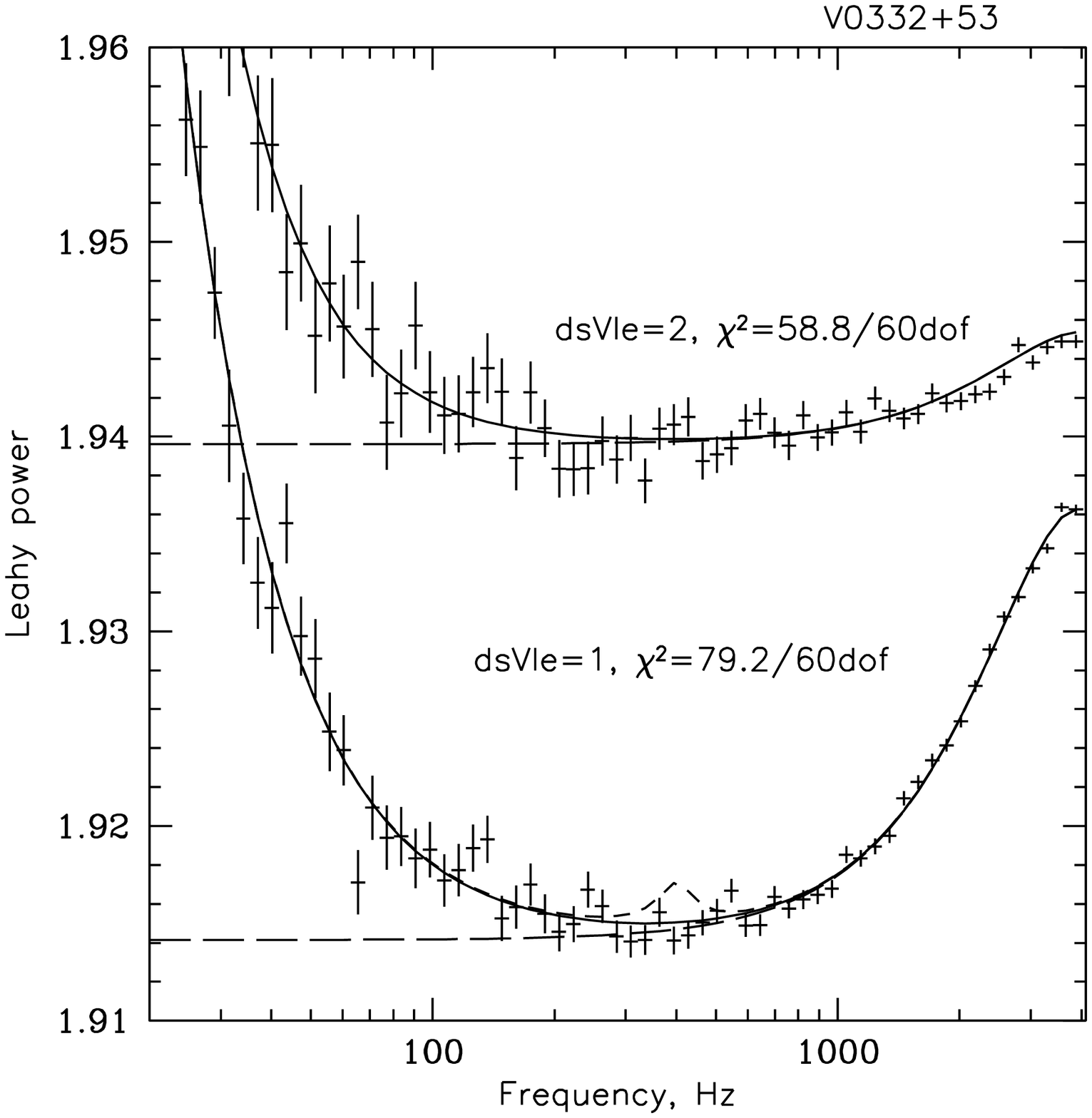}
\caption{Power spectra of V0332+53, collected during different VLE deadtime settings. Solid curves show best fit models to measured power density spectra. Long dashed curve - model of deadtime-modified Poisson noise only. Short dashed curve shows addition of Lorentzian with amplitude 0.5\%}
\label{power_v0332}
\end{figure}

In total it gives us approximately 113 ksec of observations with VLE deadtime setting {\tt dsVle=1} (average count rate $\sim$7.2 kcnts/sec/PCA) and approximately 53 ksec with VLE deadtime setting {\tt dsVle=2} (average count rate $\sim$6.2 kcnts/sec/PCA). The resulted power spectra are shown in Fig.\ref{power_v0332}.

We have fitted the obtained power spectra in frequency range 20-4096 Hz with a model, consisting of deadtime-modified Poisson noise (formulas (\ref{eqn2})) and a power law. We see that the quality of fits is good, $\chi^2$ values are 58.8 and 79.2 for 60 degrees of freedom for power spectra with VLE deadtime settings {\tt dsVle=2} and {\tt dsVle=1} respectively. During fits we have fixed the VLE deadtime at values $150~\mu$sec ({\tt dsVle=2}) and $76~\mu$sec ({\tt dsVle=1}). 

We clearly see a power law up to the Fourier frequencies 100-200 Hz (similar to \citealt{jernigan00}), but no quasiperiodic oscillations or other components, which are different from instrument-modified Poisson noise.
The $2\sigma$ upper limit on the Lorentzian QPO with quality factor 4 (i.e. $f/\Delta f=4$) is approximately 0.4-0.75\% over the Fourier frequency range 200-1500 Hz. 

Absence of some additional noise component (apart from a power law continuing from lower Fourier frequencies) at Fourier frequencies below 1-2 kHz allows us to make some conclusions about properties of the accretion column. According to calculations of \cite{klein96a,jernigan00} photon bubble oscillations noise appears below 1 kHz timescales only if polars caps have relatively large fractional area, $>10^{-3}$ of the total area of the neutron star surface. If the fractional area of the accretion column is smaller than that the typical frequencies of photon bubble oscillations were predicted to be above kHz. 

Therefore, in the framework of calculations of \cite{klein96a} we can conclude that the fractional area of the accretion column footprint on the NS surface is smaller than $<10^{-3}$. It is interesting to mention that indications that the area of the accretion column footprint on NS surfaces can be as small as $<10^{-6}$ were demonstrated in work of \cite{semena14} from completely different physical effect. Basing on upper limit on the plasma cooling time in the accretion column near magnetic white dwarf it was shown that area of the accretion column footprint is smaller than $10^{-3}$ of the WD surface. Such a small fractional area is a result of small geometrical thickness of accretion flow on top of the magnetospheric surface. Extrapolating the matter flow lines (as lines of force of a dipole magnetic field) to typical radii of NSs authors obtained the abovementioned estimate.

\section{Summary}

We have studied the power density spectra of bright X-ray sources, measured with RXTE/PCA with the aim to obtain the best constrains on presence of fast ($\sim$ kHz) noise components in lightcurves of accretion columns on magnetized neutron stars. 

\begin{itemize}
\item We have demonstrated that separation of RXTE/PCA counts into sub-bands creates distortions in their noise properties. We showed that behavior of sub-band lightcurves can be explained if all events, registered by RXTE/PCA create a sudden slight drop in detector gain, which redistribute counts, registered by PCA between energy sub-bands. The soft sub-band gain additional counts, harder energy sub-bands loose counts. 
\item Crosscorrelations between lightcurves in different energy sub-bands show that this gain returns to the original state at gaussian time scale $\sim170~\mu$sec. One of possible explanations of this effect is a screening of the electric field in the detector volume by slowly moving ions created by previous event. 
\item We have simulated this effect and obtained reasonable agreement with observed behavior of sub-bands power spectra of lightcurves of sources Sco X-1 and Cen X-3.
\item We have calculated power density spectra of one of the brightest accretion powered magnetized neutron star (pulsar) V0332+53. Avoiding the abovementioned problem with the PCA energy sub-bands we have used lightcurves of the source only in the total PCA energy bandpass. We have not detected any quasiperiodic oscillations in the Fourier frequency range 200-1500 Hz with upper limits 0.4-0.5\%.
\item In the framework of photon bubble oscillations in accretion column of magnetized neutron stars, their absence at frequencies below 1 kHz might indicate that the fractional area of footprints of the accretion column on the NS surface is small, below $10^{-3}$.
\end{itemize}

\section*{Acknowledgements}
Work is supported by grant of Russian Science Foundation 14-12-01287

\label{lastpage}
\end{document}